# Hopf-chain networks evolved from triple points

Yuee Xie,[1] Jin Cai,[1] Jinwoong Kim,[2] Po-Yao Chang,[2,3] Yuanping Chen[1]

[1]*Faculty of Science, Jiangsu University, Zhenjiang, 212013, Jiangsu, China*
[2]*Department of Physics and Astronomy, Rutgers University, Piscataway, New Jersey 08854-8019, USA*
[3]*Department of Physics, National Tsing Hua University, Hsinchu 30013, Taiwan*

## Abstract

Exotic links and chains attract interests across various disciplines including mathematics, biology, chemistry and physics. Here, we propose that topological Hopf chain networks, consisting of one-, two- and three-dimensional (3D) Hopf chains, can be found in the momentum space. These networks can be evolved from a 3D triple points phase by varying symmetries. Moreover, we identify that the Hopf-chain networks exist in a family of crystals $Sc_3XC$ ($X$ = Al, Ga, In, Tl). The crystals are 3D triple-points metals, and transit to topological metals with Hopf-chain networks under strains. These novel Hopf networks will exhibit unique magneto-transport properties.

[1] Corresponding email: chenyp@xtu.edu.cn

# I. Introduction

Zero-dimensional (0D) Hopf link, consisting of two rings linked together exactly once (see Fig. 1(a)), is the simplest nontrivial link with more than one component, according to mathematical knot theory[1,2]. Hopf links can be further periodically linked to a one-dimensional (1D) Hopf chain in Fig. 1(b), whose unit length is labeled as the light-blue dashed line[3]. A complicated 1D Hopf chain with a unit of two linked Hopf links is shown in Fig. 1(c).

Hopf links and chains have captured attentions of not only mathematicians but also biologists, chemists and physicists[4-12]. In biology, Hopf links have been found in DNA and synthesized in template synthesis, which are believed to be functionally advantageous and provide extra stability to protein chains[4-6]. In chemistry, Hopf links are well known in catenanes, a kind of compounds that consist of molecular rings held together by mechanical bonds[7-9].

In physics, the studies of topological semimetals/metals (TMs) indicated that Fermi surfaces, crossed by conduction and valence bands in the momentum space, can exhibit diverse patterns[13-53], such as points, lines and surfaces. Weyl, Dirac and triple points are representatives of topological nodal points[13-26]. Comparing with nodal points, nodal lines are more flexible to form diverse topologically distinct objects[27-50], such as rings[27-36], chains[37-40], knots[41,42] and nets[43,44]. A Hopf link, as a kind of typical topological phase including two rings[45-50], has been proposed. It shows unique topological characteristics and transport properties different from isolated and intersecting rings. All the 0D Hopf links are based on two-band models, i.e., the nodal rings are crossings of two bands. A four-band model is proposed to obtain Hopf link/chain where the two hooked nodal rings are crossings of two bands, respectively[45]. A three-band model is also proposed to get a 1D Hopf chain which is related to triple points[47].

Mathematically, 1D Hopf chains can be used to construct two-dimensional (2D) and three-dimensional (3D) networks, named as Hopf-chain networks, by linkages between chains. Figures 1(d-e) present two 2D networks consisting of two mutually vertical chains along X and Y axes, where the intersections of the chains locate at the ordinate origin. The light-blue dashed boxes show units of the networks. The difference between the two networks is that the two vertical chains in Fig. 1(e) share a blue ring while those in Fig. 1(d) intersect each other. Figure 1(f) presents a 3D network, which consists of a 2D network like that in Fig. 1(e) and a 1D simple chain along Z direction. Moreover, the 1D chain crosses over the 2D network without any contacts. Obviously, more 2D and 3D Hopf-chain networks can be obtained by different linkages between Hopf links and chains. We use a parameter $D_l^r$ to label different chains and networks, where $D$ represents dimensions, $r$ represents the number of rings in a unit, while $l$ is linking number between rings in the unit[50]. In this case, the parameters for Figs. 1(b-f) are $1_1^2$, $1_3^4$, $2_2^4$, $2_2^3$ and $3_3^5$, respectively. To date, the 2D and 3D Hopf-chain networks have not been found in momentum space.

Here, we propose that the Hopf-chain networks can be evolved from spinless triplepoints systems with multiple four-fold ($C_4$) rotation and mirror symmetries (all triple points in this work refer to spinless triple points unless otherwise specified). For example, when a system has three mutually vertical $C_4$ rotation axes and three mutually vertical mirrors, a topological phase including 7 triple points can be obtained, as shown in Fig. 3(a), based on a four-band k·p model. It evolves into a series of 3D topological networks when the structural symmetry is reduced. The evolved networks include the Hopf-chain networks in Fig. 1, as shown in Figs. 3(b-d). Moreover, we identify that these networks can be observed in the crystals $Sc_3XC$ ($X$ = Al, Ga, In, Tl) under uniaxial or biaxial strains. The symmetry requirements for the appearance of different networks are analyzed. In addition, Landau levels and topological surface states of the networks are discussed.

## II Triple points and Hopf chain networks and their k · p models

We start from a simplest Hopf chain evolved from triple points in the momentum space[47]. In a system with a 4-fold rotation axis (along $k_z$) and two mirror planes ($k_x = 0$ and $k_y = 0$), a triple-points phase can be obtained, which is shown in Fig. 2(a) and its corresponding band structures are shown in Fig. 2(d). The phase is generated by three bands. Two of the bands are degenerate along $k_z$, and they cross linearly with another band, which results in two triple points T. The degenerate band is in fact a nodal line. Each point on the nodal line is a crossing point of two quadratic bands along $k_x$ (or $k_y$), and thus the Berry phase around the point is $2\pi$[47]. Thus, the two triple points are linked by a red nodal line (see Fig. 2(a)).

When the $C_4$ rotation symmetry is eliminated but the two mirror planes are held on, the triple points evolve into a 1D Hopf chain in Fig. 2(b) or 2(c). Figures 2(e) and 2(f) present band structures for the two cases of Hopf chains, respectively. The reduced symmetry leads to the lift of the degenerate band, meanwhile the triple points T split to two crossing points $W_1$ and $W_2$. The two points $W_1$ and $W_2$ locate on two mutually vertical rings of the Hopf chain, respectively. The two rings are protected by the mirror planes $k_x = 0$ and $k_y = 0$. In the two cases of the Hopf chains, the locations of the nodal rings are different: the middle ring in Fig. 2(b) is on the plane $k_y = 0$, while that in Fig. 2(c) is on the plane $k_x = 0$. The two cases are related to the mirror eigenvalues of the two splitting bands. When the mirror eigenvalues of the two bands for point $W_2$ are (X-,

Y+) and (X-, Y-) (X± and Y± represent the eigenvalues of mirror operators on the planes $k_x = 0$, $k_y = 0$ being ±1), the middle nodal ring locates on the plane $k_y = 0$ because the two mirror eigenstates are orthogonal to each other on this mirror plane (see Fig. 2(e)). Otherwise, the middle ring locates on the plane $k_x = 0$ when the mirror eigenvalues of the two bands for point $W_2$ are (X-, Y-) and (X+, Y-) (see Fig. 2(f)). All the nodal rings in the Hopf chains have a nontrivial Berry phase π. It is noted that the Hopf chain evolved from triple points is very robust, comparing with the Hopf chain based on a four-band model [45]. The band shifting induced by perturbations may lead to disappearance of the latter, for example, the Hopf chain will change to two isolated rings. However, the former will always exist.

To obtain 2D and 3D Hopf-chain networks, a four-band k·p model with three normal 4-fold rotation axes along $k_x$, $k_y$, $k_z$ and three mirror planes on $k_x = 0$, $k_y = 0$, $k_z = 0$ is required. Considering the symmetry, the system should also have three equivalent orbitals along x, y and z, respectively. In this sense, $p_x$, $p_y$, $p_z$ and $d_{xy}$, $d_{yz}$, $d_{zx}$ can be candidates. Here, we use three $d$ orbital bands and one $s$ orbital band to construct a k·p model written as [54]

$$H(\mathbf{k}) = \begin{bmatrix} f_1 & Ak_xk_y & Bk_xk_z & Ck_yk_z \\ Ak_xk_y & f_2 & Dk_yk_z & Ek_xk_z \\ Bk_xk_z & Dk_yk_z & f_3 & Fk_xk_y \\ Ck_yk_z & Ek_xk_z & Fk_xk_y & f_4 \end{bmatrix} \quad (1)$$

where $f_i = a_i + b_i k_x^2 + c_i k_y^2 + d_i k_z^2$. Because of the three normal $C_4$ rotational symmetry, the parameters should satisfy $a_1 = a_2 = a_3$, $b_1 = c_2 = d_3$, $b_2 = b_3 = c_1 = c_3 = d_1 = d_2$, $b_4 = c_4 = d_4$, $A = B = D$, $C = E = F$.

By tuning the parameters, we obtain a phase including 7 triple points, as shown in Fig. 3(a). There are a pair of triple points on each axis, and there is another triple point at the center Γ of Brillouin zone (BZ). The band structures along three axes are given in Fig. 3(e). Beside the top parabolic band, the band structures are similar to that in Fig. 2(d): a degenerate band crosses linearly with the other single band, which generates six triple points $T_1$ on $k_{x/y/z}$ axis, and these triple points are linked by straight nodal lines along axes. The top parabolic band in Fig. 3(e) introduces additional triple point $T_2$ at Γ point which quadratically touches the degenerate band. Therefore, the phase in Fig. 3(a) is a 3D version of the 1D triple points phase in Fig. 2(a) adding an extra triple point at Γ point, i.e., a 3D triple-points phase.

When one $C_4$ rotation symmetry, for example, the $C_4$ symmetry along $k_y$, is broken, another $C_4$ rotation symmetry will also be broken, say $k_x$. Then, only the $C_4$ rotation symmetry along $k_z$ is held on. Different values of the parameters will result in two different topological phases, as shown in Figs.3(b) and 3(c), respectively. The band structure corresponding to the phase in Fig.3(b) is shown in Fig. 3(f). The degenerate band along $k_x$ and $k_y$ axes is lifted by the broken $C_4$ rotation symmetries, and the triple point $T_1$ on $k_{x/y}$ axis splits to two crossing points. This induces the transition on $k_x$ and $k_y$ axes from triple points to Hopf chain like that from Fig. 2(a) to 2(c). Seen from the mirror eigenvalues of the crossing bands, the central rings of the mutually orthogonal chains locate on the planes $k_x = 0$ and $k_y = 0$, respectively, and they contact each other on the $k_z$ axis forming intersecting rings. As a result, the 2D Hopf-chain network discussed in Fig. 1(d) is materialized as shown in Fig. 3(b). Along $k_z$ axis, the two triple points still exist because of the $C_4$ rotation symmetry is reserved. If we consider the top band, the lift of the degenerate band leads to disappearance of the triple point $T_2$ at Γ point. The triple point $T_2$ evolves into a nodal ring with center at Γ point on the plane $k_z$

= 0. The whole topological phase is shown in Fig. 3(b), which is a mixture of a 2D Hopf-chain network, a nodal ring and two triple points.

Another phase resulted from breaking two $C_4$ rotation symmetries is shown in Fig. 3(c), and its corresponding band structure is given in Fig. 3(g). By comparing Figs. 3(f) and 3(g), one can find that the brown band shifts down in Fig. 3(g). The shifting exchanges the mirror eigenvalues of bands, which results in the central nodal rings of the two chains both locate on the plane $k_z = 0$ instead of the two vertical mirror planes. The two rings combine together to a distorted ring which is shared by the two Hopf chains. Obviously, the two orthogonal chains form a 2D Hopf-chain network similar to that in Fig. 1(e). Another change induced by the shifting band is the original triple point at Γ point in Fig. 3(a) transits to two triple points along $k_z$ axis. Therefore, the topological phase in Fig. 3(c) shows a mixture of a 2D Hopf-chain network and four triple points on $k_z$ axis.

Next, we consider the case that all the three $C_4$ rotation symmetries are broken while the three mirror planes are still held on. In this case, the constraints to all parameters in Eq. (1) are released. Figure 3(d) shows a topological phase evolved from Fig. 3(c) (or Fig. 3(b)), and the corresponding band structure is given in Fig. 3(h). Because all the $C_4$ rotation symmetries are eliminated, the degenerate bands along all axes split, and thus the two couples of triple points on $k_z$ axis in Fig. 3(C) split to a complicated 1D Hopf chain along $k_z$ axis. The emergent Hopf chain along $k_z$ axis is similar to that in Fig. 1(c). This Hopf chain and the 2D Hopf chain along $k_x$ and $k_y$ axes form a 3D Hopf-chain network. The detailed values of the parameters for all phases are shown in Table 1. By tuning the parameters in Eq. (1), some other new topological network phases are given in Fig. 7 in Appendix A and their corresponding parameters are listed in Table 1 in Appendix A.

## III Realization of topological phases in real crystals

In order to realize these topological phases in real crystals, the crystal structures should have three $C_4$ rotation symmetries along x, y and z axes and mirror planes on the three coordinate planes. We find that the family of Sc$_3$XC (X= Al, Ga, In, Tl), which crystallizes in cubic structure with space group PM-3M(OH-1), can meet the requirement. Their atomic structures are schematically shown in Fig. 4(a). The element C is located at the center and the atoms Sc and X are located at the face center and the cubic corner, respectively[55,56]. Structure parameters of the crystals are given in Table 2 in Appendix B.

We performed first-principles calculations within the density functional theory (DFT) as implemented in the VASP codes[57,58]. The potential of the core electrons and the exchange-correlation interaction between the valence electrons were described, respectively, by the projector augmented wave[59] and the generalized gradient approximation (GGA) with Perdew-Burke-Ernzerhof (PBE)[60] functional. A kinetic energy cutoff of 550 eV was used. The atomic positions were optimized using the conjugate gradient method, and the energy and force convergence criteria were set to be $10^{-5}$ eV and $10^{-3}$ eV/Å, respectively. The BZ was sampled in the k space within the Monkhorst-Pack scheme and the number of these k points was 7×7×7. The surface energy dispersions were calculated within the tight binding scheme based on the maximally localized Wannier functions (MLWFs), using the Wannier-tools software package[61].

We calculate band structure of the family of $Sc_3XC$ ($X=$ Al, Ga, In, Tl), and the results indicate that the crystals have very similar band structures (see Fig. 8 in Appendix B). Figure 4(c) show the band spectrum of $Sc_3InC$ in the case of no spin-orbit coupling (SOC). When SOC is absent, the system can be considered as a spinless system. One can find that the band spectrum along $\Gamma$-X is very similar to that in Fig. 2(a): there is a triple point on $k_x$ axis and a triple point at the $\Gamma$ point. Therefore, these structures possess the 3D triple-points phase in Fig. 3(a). By applying different strains, the topological phase evolves into various other phases which have been obtained by the k · p model. For example, when a tensile or a compressive strain is applied on z axis to eliminate the two $C_4$ rotation symmetries along x and y axes, mixture phases in Figs. 3(b-c) can be obtained, respectively. When a strain is further applied on x or y axis to eliminate the final $C_4$ rotation symmetry, the phase evolves to a 3D Hopf-chain network like that in Fig. 3(d). Band structures corresponding to the topological phases as well as contour plots for energy differences of bands are given in Fig. 9 in Appendix B.

If SOC is considered, the original triple points and nodal rings are gapped (see Fig. 11 in Appendix B and insets in Fig. 4(c)). Each band is still doubly degenerate because of the presence of inversion symmetry. Because the gaps of nodal rings are very small (only 1 ~ 6 meV), as shown in Fig. 11 in Appendix B, the Hopf-chain networks discussed above can be observed experimentally.

The Hopf chain consists of mutually orthogonal rings. The rings can be projected on the mutually vertical surfaces and lead to drumhead states respectively. The calculations of surface states for $Sc_3InC$ identify existence of the drumhead states, as shown in Fig. 5. Figures 5($a_1$-$d_1$) present bulk band structures for different topological phases, while the corresponding surface band structures are shown in Figs. ($a_2$-$d_3$). On the surfaces of the 3D triple-points phase, there are "tent" states between the triple points [see Figs. 5(a$_2$-a$_3$)]. These surface states evolve into drumhead states after the triple points evolve into Hopf chains. For example, for the 2D Hopf-chain phases in Figs. 3(b) and 3(c), the nodal rings are projected on the surfaces [100] and [001], respectively, and thus drumhead states are found on the two surfaces [see Figs. 5(b$_2$-c$_3$)]; moreover, the distributions of the nodal rings in the two phases are inverse, which results in inverse surface states, i.e., the surface state on the surface [100] for the phase in Fig. 3(b) corresponds to that on the surface [001] for the phase in Fig. 3(c). Because the nodal rings along $k_z$ axis in the 3D Hopf-chain networks are too small (see Fig. 9(d$_5$)), the drumhead states in them cannot be distinguished and thus the surfaces states in Figs. 5(d$_2$-d$_3$) seem similar to those in Figs. 5(c$_2$-c$_3$).

## IV Landau level spectra of the topological phases

We calculate the Landau levels of the four phases in Figs. 3(a-d) based on the k·p model in Eq (1) [62]. We consider the case the external magnetic field ***B*** is applied along z-direction. Since ***B*** is along z-direction, the k·p Hamiltonian can be parametrized by $k_z$ and two conjugate momenta, $\Pi_x = k_x - eA_x = \frac{1}{\sqrt{2}l_B}(a^+ + a)$ and $\Pi_y = k_y - eA_y = \frac{1}{\sqrt{2}il_B}(a^+ - a)$. Here $l_B = 1/\sqrt{eB}$ is the magnetic length, $a^+$ and $a$ are the raising and lowering operators respectively. We can solve the eigenequation $H(k_z, B)\psi = E(k_z, B)\psi$ with the eigenfunction $\psi = \sum_n(\alpha_n|n\rangle, \beta_n|n\rangle, \gamma_n|n\rangle)^T$. $|n\rangle$ satisfies $a^+|n\rangle = \sqrt{n+1}|n+1\rangle$ and $a|n\rangle = \sqrt{n}|n-1\rangle$. Then the eigenequation becomes coupled equations. With proper truncation of the Landau index *n*, the Landau level spectrum is obtained.

As shown in Fig. 6, one can find that there are two kinds of Landau levels in the spectra. One is chiral modes induced by the triple points [63], the other is zeroth Landau modes induced by the nodal rings [62]. In the topological phases in Fig. 3(a-c), there are triple points along $k_z$ axis, and thus chiral modes appear on the spectra in Fig. 6(a-c). Especially in the phase of Fig. 3(c), there are two pairs of triples points, correspondingly, one can find two sets of chiral modes between them. After all the triple points split to nodal rings, in Fig. 6(d), zeroth Landau modes appear and are bounded by the projection of nodal rings. To conclude, while two triple points splits and evolves to the Hopf link, the original chiral modes connected two triple points in the Landau level spectrum becomes a zeroth Landau mode located within the inner projection of the nodal rings.

## V Discussions and Summary

In the real materials Sc$_3$XC, the triple points and the Hopf-chain networks are not exactly located at the Fermi level. Therefore, many electron and hole pockets exist near the Fermi level, which results in semimetallic behavior in terms of transports. Under the low magnetic field, these materials are expected to have extreme magnetoresistance due to the compensation between the electron and hole carriers[64-66]. In the quantum region, when the quantized Landau orbits lead to quantum oscillations, one can expect the magnetic breakdown due to the tunneling between electron and hole pockets. The magnetic breakdown are strongly dependent on the direction of the magnetic field[67]. In the case of Hopf-chain network extending on the $k_x$-$k_y$ plane (see Fig. 1(e)), one can expect strong magnetic breakdown when the field is along the $k_z$ direction. On the other hand, in the case of Hopf-chain network extending on the entire Brillouin zone (see Fig. 1(f)), profound magnetic breakdown can be observed in all three axes.

The Hopf-chain networks discussed here are related to a four-band case. One can expect that more abundant Hopf-chain networks can be found in multi-band cases. These exotic Hopf patterns are common interests of mathematics, biology, chemistry and physics. Therefore, our work not only provides new topological phases and emergent fermions in physics, but also opens a door for other fields to explore the beautiful motifs.

## Acknowledgments

We thank the discussions with Heung-Sik Kim. This work was supported by the National Science Foundation of China (No. 11474243 and No. 51376005).

**Appendix A: k · p model for the triple-points phase and Hopf-chain networks**

Let us consider a system with three $C_4$ rotation symmetries along three cartesian axes and three mirrors on the three coordinate planes. Four atomic orbits, $\{d_{yz}, d_{xz}, d_{xy}, s\}$ are considered as basis. Then, the symmetry operators for the $C_4$ rotation symmetries along cartesian axes can be written as

$$C_{4x} = \begin{bmatrix} -1 & 0 & 0 & 0 \\ 0 & 0 & -1 & 0 \\ 0 & 1 & 0 & 0 \\ 0 & 0 & 0 & 1 \end{bmatrix}, \quad C_{4y} = \begin{bmatrix} 0 & 0 & 1 & 0 \\ 0 & -1 & 0 & 0 \\ -1 & 0 & 0 & 0 \\ 0 & 0 & 0 & 1 \end{bmatrix}, \quad C_{4\hat{z}} = \begin{bmatrix} 0 & -1 & 0 & 0 \\ 1 & 0 & 0 & 0 \\ 0 & 0 & -1 & 0 \\ 0 & 0 & 0 & 1 \end{bmatrix}, (2)$$

while operators for the mirror symmetries can be written as

$$M_x = \begin{pmatrix} 1 & 0 & 0 & 0 \\ 0 & -1 & 0 & 0 \\ 0 & 0 & -1 & 0 \\ 0 & 0 & 0 & 1 \end{pmatrix}, \quad M_y = \begin{pmatrix} -1 & 0 & 0 & 0 \\ 0 & 1 & 0 & 0 \\ 0 & 0 & -1 & 0 \\ 0 & 0 & 0 & 1 \end{pmatrix}, \quad M_z = \begin{pmatrix} -1 & 0 & 0 & 0 \\ 0 & -1 & 0 & 0 \\ 0 & 0 & 1 & 0 \\ 0 & 0 & 0 & 1 \end{pmatrix}. \quad (3)$$

If the system has both inversion and TR symmetries, the Hamiltonian $H(k_x,k_y,k_z)$ is real. Each matrix element of the Hamiltonian is expanded up to harmonic order in the wave vectors $k_i$ and higher order terms are truncated without loss of generality in the vicinity of Γ point. Since the mirror and two-fold rotation symmetries are preserved even after the breakdown of the $C_4$ rotation symmetries, those symmetries, $M_i$ and $C_{2i}$, are first considered as least constraints to the Hamiltonian. Applying the least constraints, we obtain

$$H(k_x, k_y, k_z) = \begin{bmatrix} f_1 & g_1 & g_2 & g_3 \\ g_1{}^* & f_2 & g_4 & g_5 \\ g_2{}^* & g_4{}^* & f_3 & g_6 \\ g_3{}^* & g_5{}^* & g_6{}^* & f_4 \end{bmatrix}, \quad (4)$$

where $f_i = a_i + b_i k_x^2 + c_i k_y^2 + d_i k_z^2$, $g_1 = Ak_xk_y$, $g_2 = Bk_xk_z$, $g_3 = Ck_yk_z$, $g_4 = Dk_yk_z$, $g_5 = Ek_xk_z$, $g_6 = Fk_xk_y$. The remaining $C_4$ rotation symmetries give additional constraints which are,

$$a_2 = a_3, b_2 = b_3, c_2 = d_3, d_2 = c_3, A = B, E = F, \text{ (with } C_{4x}) \quad (5)$$
$$a_1 = a_3, b_1 = d_3, c_1 = c_3, d_1 = b_3, A = D, C = F, \text{ (with } C_{4y}) \quad (6)$$
$$a_1 = a_2, c_1 = b_2, b_1 = c_2, d_1 = d_2, B = D, C = E. \text{ (with } C_{4\hat{z}}) \quad (7)$$

Considering all the symmetry constraints, we eventually get the k·p model Hamiltonian with the maximum symmetry,

$$H(\mathbf{k}) = \begin{bmatrix} f_1 & Ak_xk_y & Bk_xk_z & Ck_yk_z \\ Ak_xk_y & f_2 & Dk_yk_z & Ek_xk_z \\ Bk_xk_z & Dk_yk_z & f_3 & Fk_xk_y \\ Ck_yk_z & Ek_xk_z & Fk_xk_y & f_4 \end{bmatrix}, \quad (8)$$

with $a_1= a_2= a_3$, $b_1= c_2= d_3$, $b_2= b_3= c_1= c_3= d_1= d_2$, $b_4= c_4= d_4$, $A= B= D$, $C= E= F$. Therefore, it is clear that three $f_i$ are equivalent copies under permutation of $\{k_x, k_y, k_z\}$ and two of them become degenerate on the cartesian axes.

For an example, three pairs of triple points are guaranteed to emerge by taking $a_1>a_4$ and $c_1<c_4$ which are crossings between $f_4$ and $f_i$ (i= 1, 2, 3) marked as T$_1$ in Fig. S3(a). Breakdown of $C_{4x}$ and $C_{4y}$ removes the constraints of Eqs. (5) and (6), and the relevant degeneracies are lifted. The proper choice of parameters may render the transition to the Hopf-chain networks as listed in Table S1.

One can also use other four atomic orbits $\{p_x, p_y, p_z, s\}$ as a basis. Then, the symmetry operators for the C$_4$ rotation symmetries along cartesian axes can be written as

$$C_{4x}=\begin{bmatrix}1 & 0 & 0 & 0\\ 0 & 0 & -1 & 0\\ 0 & 1 & 0 & 0\\ 0 & 0 & 0 & 1\end{bmatrix}, \quad C_{4y}=\begin{bmatrix}0 & 0 & 1 & 0\\ 0 & 1 & 0 & 0\\ -1 & 0 & 0 & 0\\ 0 & 0 & 0 & 1\end{bmatrix}, \quad C_{4\hat{z}}=\begin{bmatrix}0 & -1 & 0 & 0\\ 1 & 0 & 0 & 0\\ 0 & 0 & 1 & 0\\ 0 & 0 & 0 & 1\end{bmatrix}, \quad (9)$$

while operators for the mirror symmetries can be written as

$$M_x=\begin{pmatrix}-1 & 0 & 0 & 0\\ 0 & 1 & 0 & 0\\ 0 & 0 & 1 & 0\\ 0 & 0 & 0 & 1\end{pmatrix}, \quad M_y=\begin{pmatrix}1 & 0 & 0 & 0\\ 0 & -1 & 0 & 0\\ 0 & 0 & 1 & 0\\ 0 & 0 & 0 & 1\end{pmatrix}, \quad M_z=\begin{pmatrix}1 & 0 & 0 & 0\\ 0 & 1 & 0 & 0\\ 0 & 0 & -1 & 0\\ 0 & 0 & 0 & 1\end{pmatrix}, \quad (10)$$

If the system has both inversion and TR symmetries, the Hamiltonian, $H(k_x,k_y,k_z)$ is real. Each matrix element of the Hamiltonian is expanded up to harmonic order in the wave vectors $k_i$ and higher order terms are truncated without loss of generality in the vicinity of Γ point. Since the mirror and two-fold rotation symmetries are preserved even after the breakdown of the C$_4$ rotation symmetries, those symmetries, $M_i$ and $C_{2i}$, are first considered as least constraints to the Hamiltonian. Applying the least constraints, we obtain

$$H(k_x, k_y, k_z) = \begin{bmatrix} f_1 & g_1 & g_2 & g_3 \\ g_1^* & f_2 & g_4 & g_5 \\ g_2^* & g_4^* & f_3 & g_6 \\ g_3^* & g_5^* & g_6^* & f_4 \end{bmatrix}, \quad (11)$$

Where $f_i = a_i + b_i k_x^2 + c_i k_y^2 + d_i k_z^2$, $g_1 = A k_x k_y + a k_z$, $g_2 = B k_x k_z + b k_y$, $g_3 = C k_y k_z + c k_x$, $g_4 = D k_y k_z + d k_x$, $g_5 = E k_x k_z + e k_y$, $g_6 = F k_x k_y + f k_z$. The remaining C4 rotation symmetries give additional constraints which are;

$$\begin{aligned} &a_2 = a_3, b_2 = b_3, c_2 = d_3, d_2 = c_3, c_4 = d_4 \\ &A = -B, E = -F, C = 0, a = b, e = f, d = 0 \end{aligned}, \quad \text{(with } C_{4x}\text{)} \quad (12)$$

$$\begin{aligned} &a_1 = a_3, b_1 = d_3, c_1 = c_3, d_1 = b_3, b_4 = d_4 \\ &A = -D, C = F, E = 0, a = d, c = -f, b = 0 \end{aligned}, \quad \text{(with } C_{4y}\text{)} \quad (13)$$

$$\begin{aligned} &a_1 = a_2, c_1 = b_2, b_1 = c_2, d_1 = d_2, b_4 = c_4 \\ &B = D, c = -e \end{aligned}, \quad \text{(with } C_{4z}\text{)} \quad (14)$$

Considering all the symmetry constraints, we eventually get the *k·p* model Hamiltonian with the maximum symmetry,

$$H(\mathbf{k}) = \begin{bmatrix} f_1 & A k_x k_y & B k_x k_z & c k_x \\ A k_x k_y & f_2 & D k_y k_z & e k_y \\ B k_x k_z & D k_y k_z & f_3 & f k_z \\ c k_x & e k_y & f k_z & f_4 \end{bmatrix}, \quad (15)$$

where $a_1 = a_2 = a_3$, $b_1 = c_2 = d_3$, $b_2 = b_3 = c_1 = c_3 = d_1 = d_2$, $b_4 = c_4 = d_4$, $A = -B = -D$, $C = -E = F = 0$, $a = b = d = 0$, $c = -e = -f$. Therefore, it is clear that three $f_i$ are equivalent copies under permutation of $\{k_x, k_y, k_z\}$ and two of them become degenerate on the cartesian axes. For an example, three pairs of triple points are guaranteed to emerge by taking $a_1 > a_4$ and $c_1 < c_4$ which are crossings between $f_4$ and $f_i$ (i= 1, 2, 3). Breakdown of $C_{4x}$ and $C_{4y}$ removes the constraints Eqs. (12) and (13) and the relevant degeneracies are lifted

$$a_1 = a_2,\ c_1 = b_2,\ b_1 = c_2,\ d_1 = d_2,\ b_4 = c_4,\ B = D,\ c = -e. \qquad (16)$$

Based on Eq. (1) (or Eqs. (8) and (15)), many topological phases can be obtained by tuning the parameters, for example, the topological phases in Figs. 3(a-d) and those in Figs. 7(a-c). The detailed values of the parameters can be found in Table 1.

**Appendix B: Band structures and structural parameters of $Sc_3X C$**

The family of $Sc_3XC$ ($X=$ Al, Ga, In, Tl) has the same crystal structure with space group PM-3M(OH-1). Their structural parameters, such as lattice constants and bond lengths, are given in Table 2. Band structures in the case of no SOC is given in Fig. 8. One can find these band structures are similar especially around the Fermi level. To clearly exhibit the topological phases and phase transitions in the real crystals, we calculate evolution of band structures of $Sc_3InC$ under different strains, and the corresponding contour plots of energy difference between two bands are shown (see Fig. 9). We also calculate Fermi surfaces in $Sc_3InC$ under different strains, as shown in Fig. 10. The effect of SOC on the band structure is considered in Fig. 11.

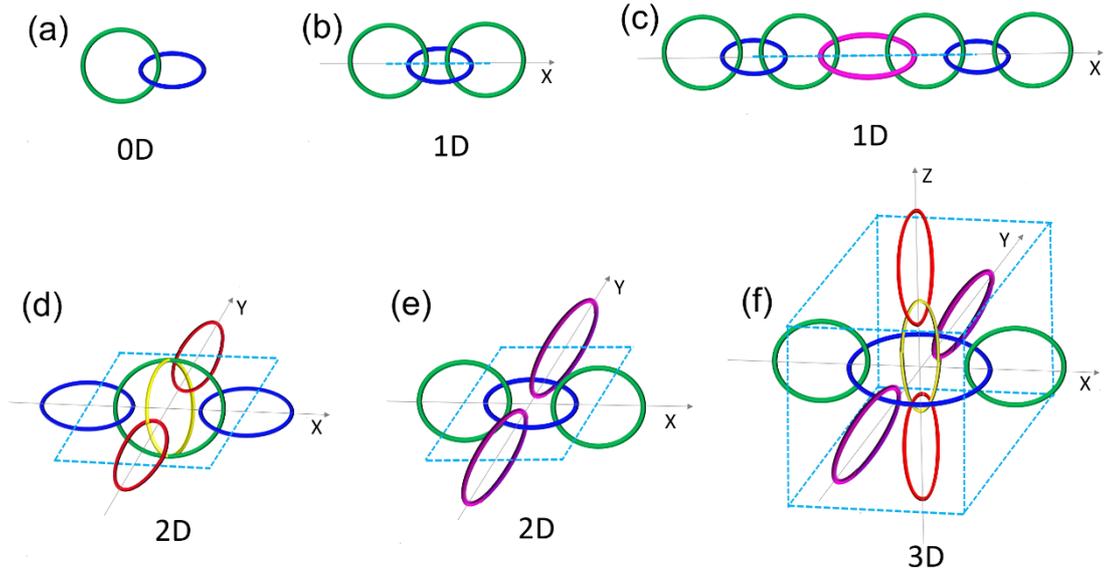

FIG.1. **Hopf link, Hopf chains and Hopf-chain networks.** (a) Hopf link. (b) 1D Hopf chain with a unit consisting of one Hopf link. (c) 1D Hopf chain with a unit consisting of two Hopf links. (d-e) 2D Hopf-chain networks formed by two mutually vertical chains along X and Y axes. The difference between the two networks is that the two vertical chains in (e) share a blue ring while those in (d) intersect each other. (f) 3D Hopf-chain network, consisting of a 2D network like that in (e) and a 1D chain like that in (b) along Z direction, where the 1D chain crosses over the 2D network without any contacts. The unit length of each periodic structure is labeled as light-blue dashed lines.

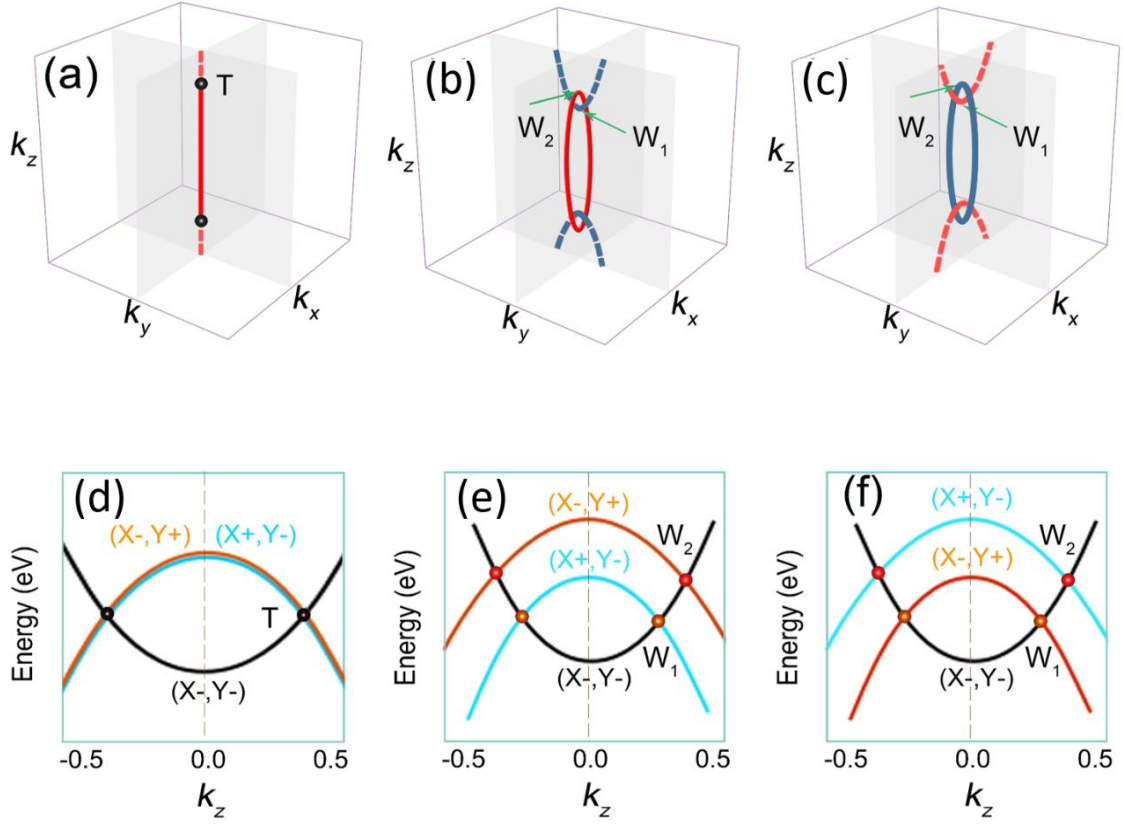

FIG.2. **Transition from a triple-points phase to a 1D Hopf chain.** (a) A triple-points phase, where two (black) triple points T are connected by a (red) straight nodal line along $k_z$. (b-c) 1D Hopf chains where two nodal rings link each other. The two mutually vertical rings locate on two mirror planes $k_x = 0$ and $k_y = 0$, respectively, and the middle ring is centered at Γ point. The difference between (b) and (c) is that the two rings lie on the different mirror planes. The middle red ring in (b) is on the mirror plane $k_y = 0$ while the middle blue ring in (c) is on the mirror plane $k_x = 0$. The phase transition from (a) to (b) or (c) is induced by the elimination of $C_4$ rotation symmetry. (d-f) Band structures corresponding to the topological phases in (a-c) along $k_z$, respectively. The double degenerate band in (d) splits into two bands in (e) or (f), which results in the triple point T in (d) splitting into two crossing points $W_1$ and $W_2$ in (e) or (f). In (e) the splitting red band is above on the light-blue band, while the case in (f) is inverse. X± and Y± in the parentheses in (d-f) represent the eigenvalues of mirror planes $k_x = 0$, $k_y = 0$ being ±1, respectively.

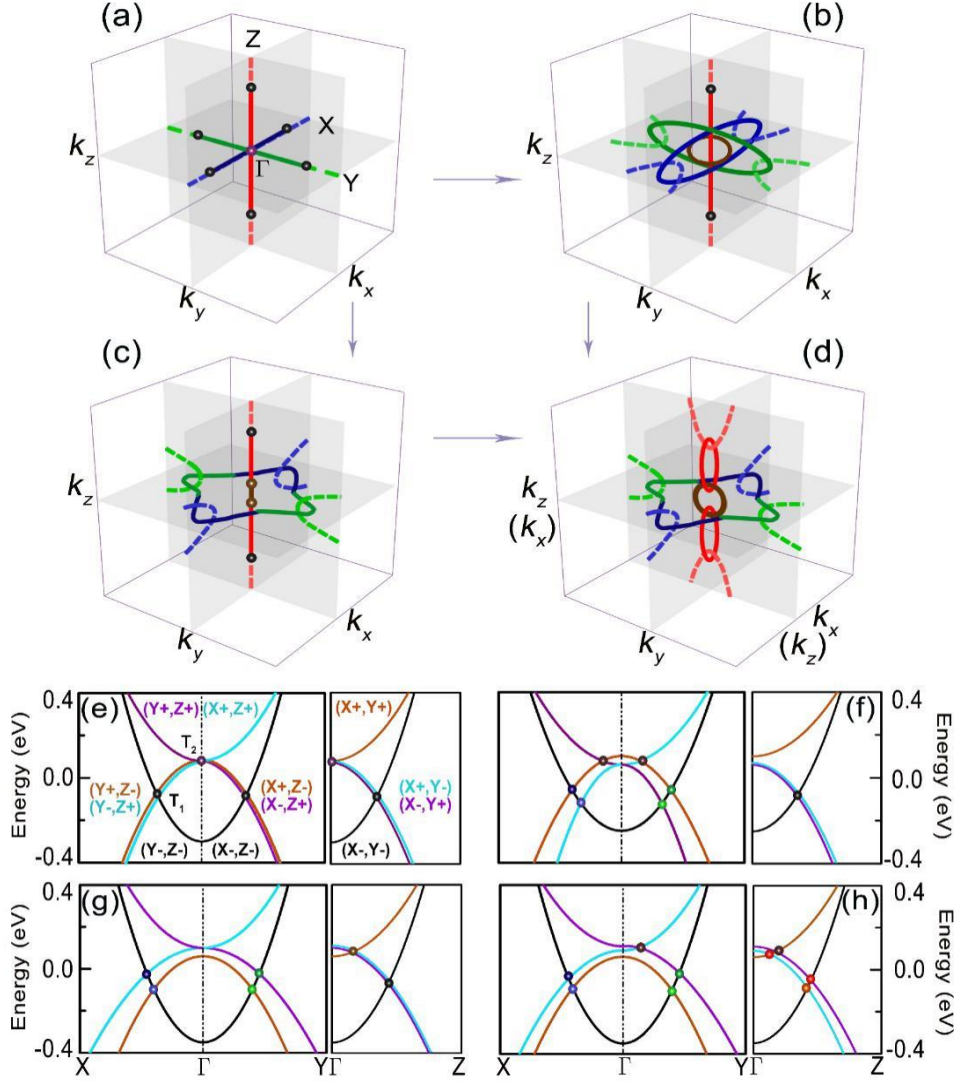

FIG.3. **Transition from a 3D triple-points phase to 2D/3D Hopf-chain networks.** (a) A 3D triple-points phase with 7 triple points, which is a 3D version of the phase in Fig. 2(a). (b) A mixture phase of a 2D Hopf-chain network (like that in Fig. 1(d)), a nodal ring and two triple points. (c) A mixture phase of a 2D Hopf-chain network (like that in Fig. 1(e)), and four triple points. (d) A 3D Hopf-chain networks. (e-h) Band structures corresponding to the phases in (a-d), respectively, in which X±, Y± and Z± in the parentheses represent the eigenvalues of mirror planes $k_x=0$, $k_y=0$ and $k_z=0$ being ±1. All the phases and corresponding band structures are generated by Eq. (1) with different parameters. The phase in (a) is protected by three normal $C_4$ rotation axes along $k_{x/y/z}$ and three mirror planes. When the $C_4$ rotation axes along $k_{x/y}$ are broken, it evolves into the phases in (b) and (c). When the $C_4$ rotation axis along $k_z$ is further broken, the phases (b) and (c) evolve into the phase in (d). The coordinate axes in and out the parentheses in (d) represent the axes for the phase transiting from (b) and (c), respectively.

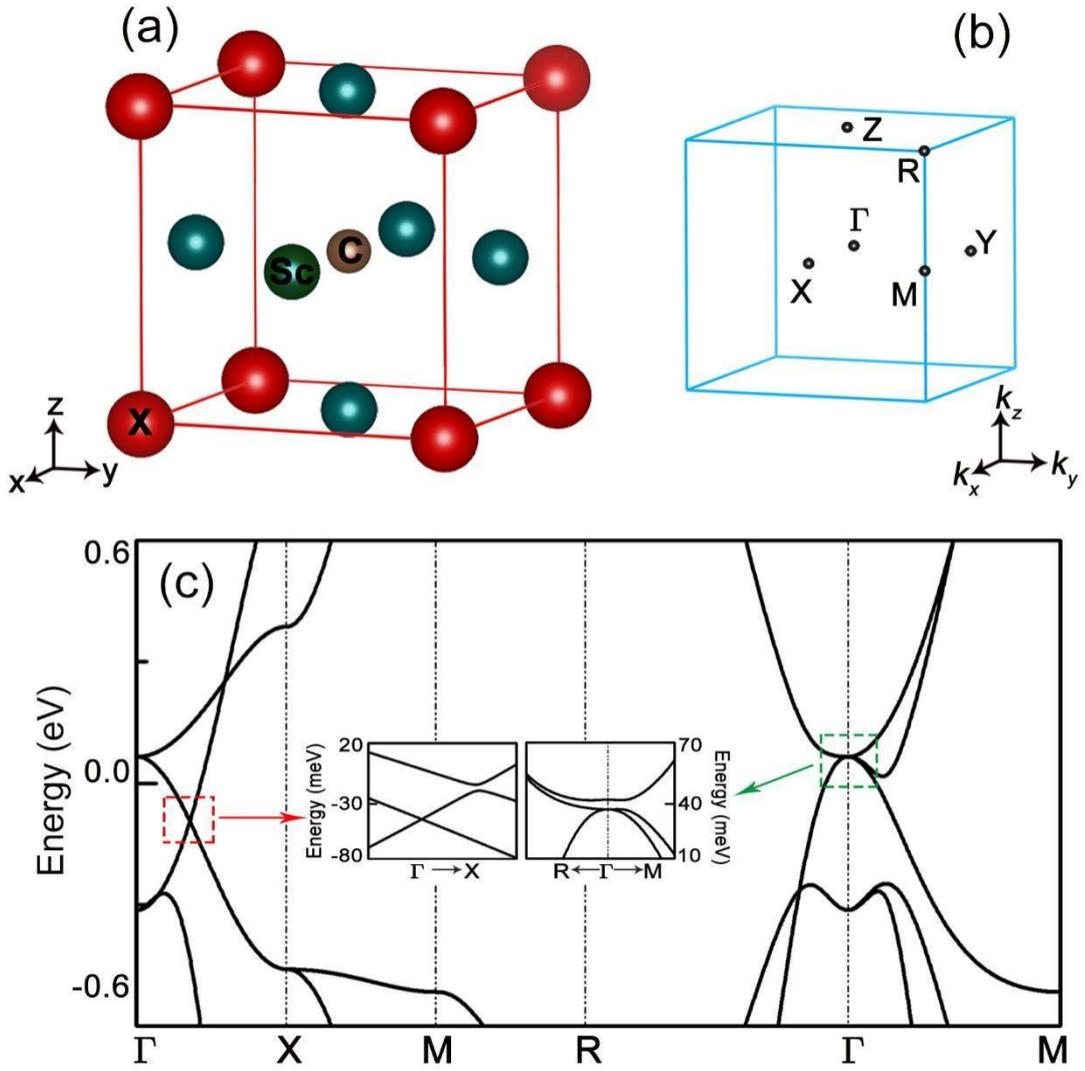

FIG. 4. **Material realizations of the triple-points and 2D/3D Hopf-chain networks in Sc₃XC.** (a) Crystal structure of Sc$_3$XC ($X$ = Al, Ga, In, Tl). (b) BZ with high symmetry points marked. (c) Band structure of Sc$_3$InC in the case of no SOC, and the insets show the band structures near the triple points in the case of considering SOC.

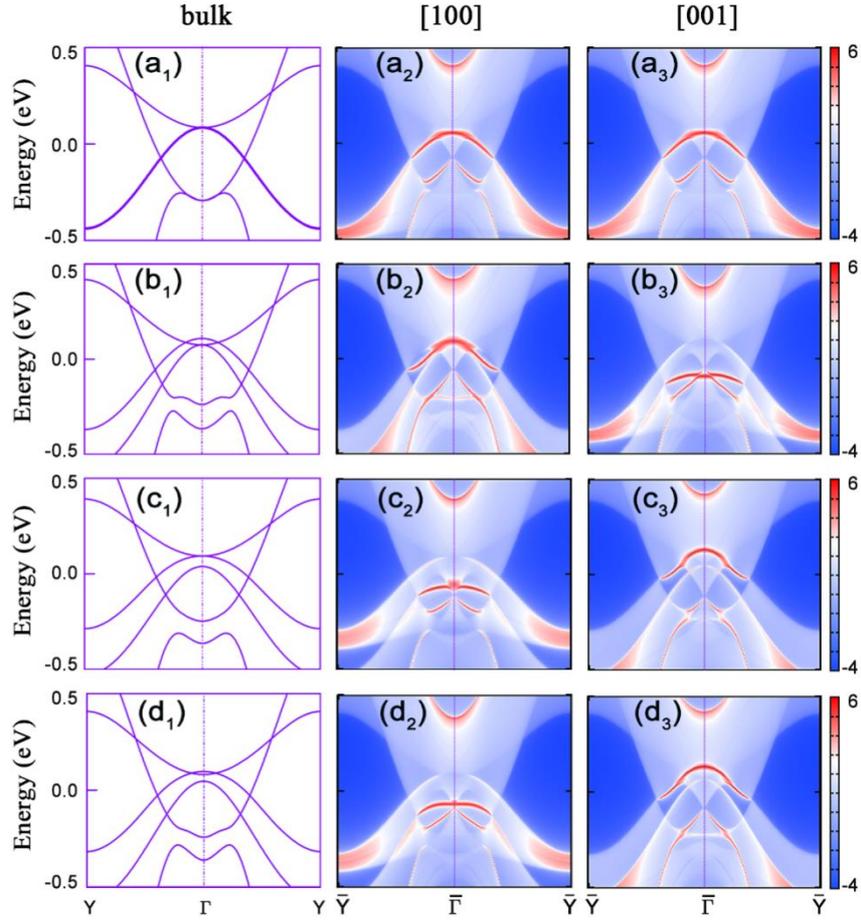

FIG. 5. Bulk band structures along Y-Γ-Y of Sc$_3$InC in cases of (a$_1$) no strain, (b$_1$) 3% (5.30 GPa) tensile strain along $z$ axis, (c$_1$) 3% (6.59 GPa) compressive strain along $z$ axis, (d$_1$) 3% (7.27 GPa) compressive strain along $z$ axis and 1% (3.21 GPa) compressive strain along $y$ axis. (a$_1$- d$_1$) are similar to Figs. 3(e-h) along the same k path, respectively, and thus the cases in (a$_1$-d$_1$) correspond to the topological phases in Figs. 3(a-d). (a$_2$-d$_2$) Surface band structures on the [100] surface corresponding to the bulk band structures in (a$_1$-d$_1$), respectively. (a$_3$-d$_3$) Surface band structures on the [001] surface corresponding to (a$_1$-d$_1$), respectively.

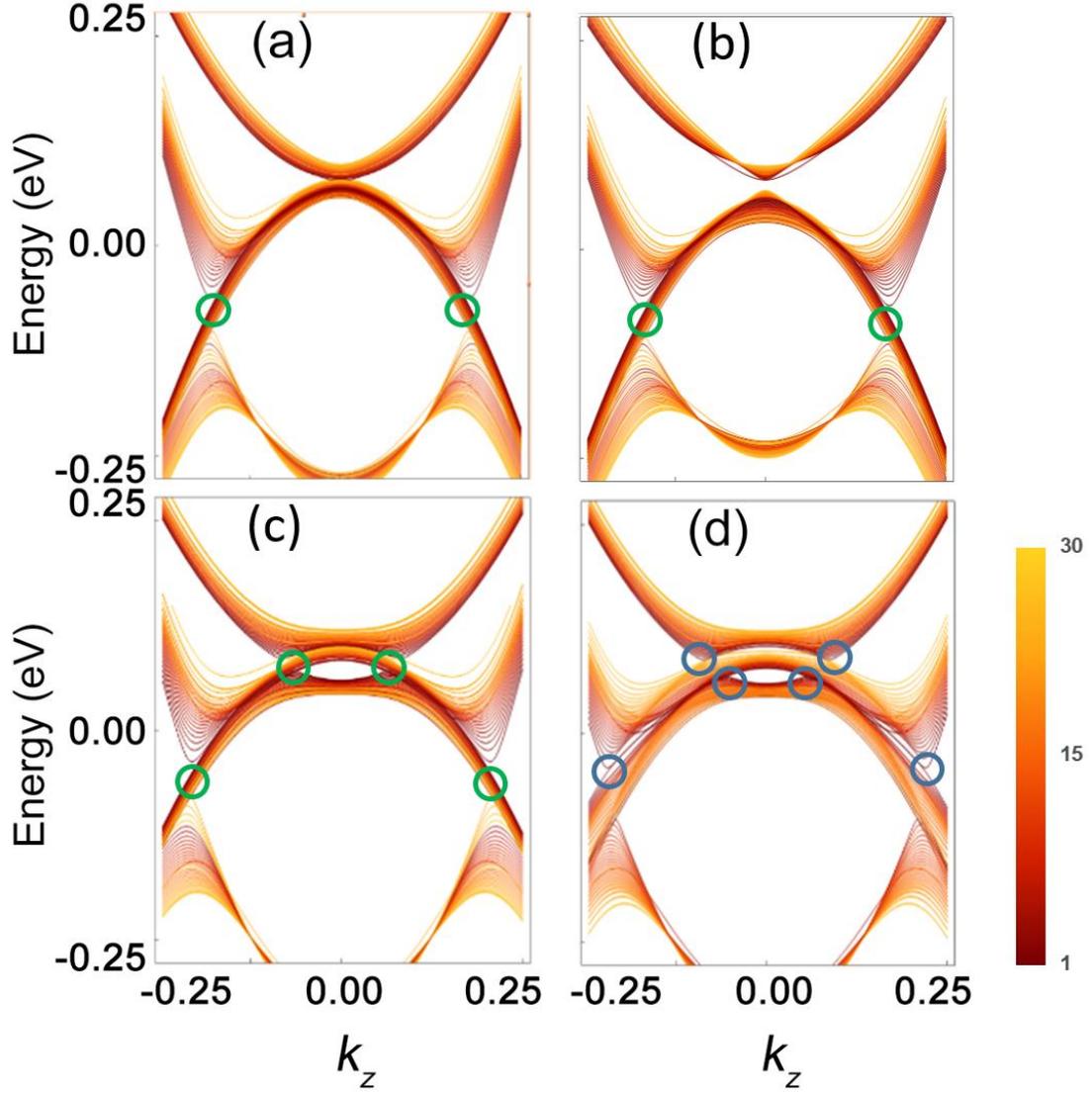

FIG. 6. (a-d) Landau level spectra for the topological phases in Figs. 3(a-d), respectively. The empty green dots represent the projected triple points while the empty blue dots represent two ends of the projected nodal rings. The magnetic field $\boldsymbol{B}$ is along $z$ axis. The color intensity is the Landau index from $n = 1$ to 30.

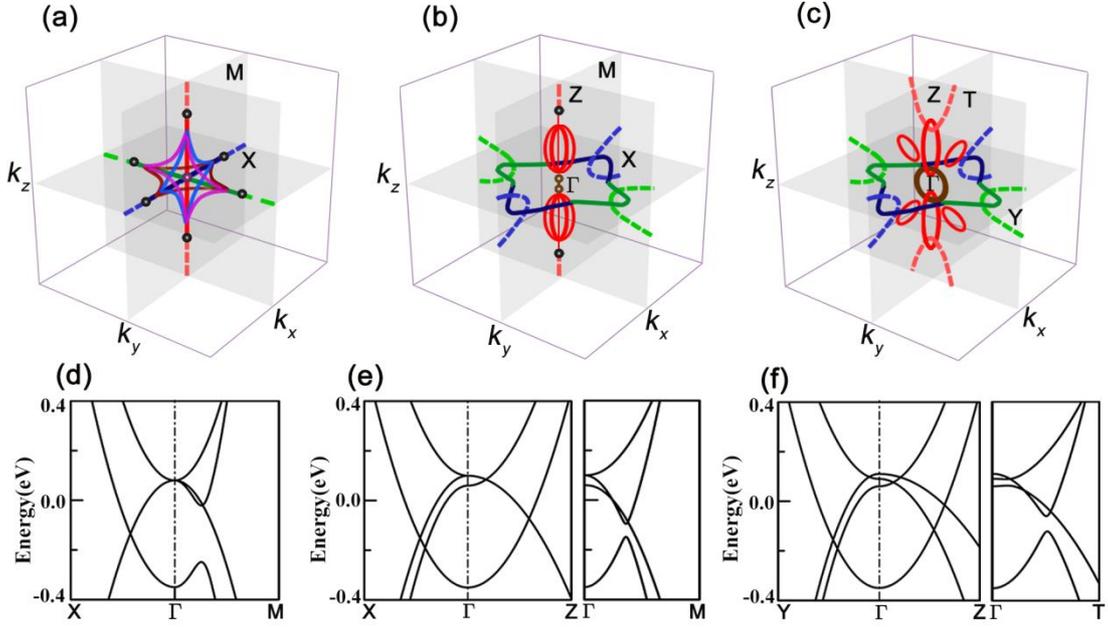

FIG. 7. Other new topological network phases generated by Eq. (1), whose parameters are listed in Table S1. (a) A mixture phase of triple points and additional nodal lines, which has the same symmetry with the phase in Fig. 3(a). (b) A phase including a 2D Hopf-chain network, four triple points and additional nodal rings. This phase is equivalent to the phase in Fig. 3(c) adding four red nodal rings. (c) A phase including a 3D Hopf-chain network and four red additional nodal rings. This phase is equivalent to the phase in Fig. 3(d) adding four red nodal rings. All the additional nodal rings and lines are located on the mirror planes $k_{x/y/z} = 0$. (d-f) Band structures corresponding to the phases in (a-c), respectively. It is noted that the phase in (a) can be found in the crystals $Sc_3XC$ ( $X$ = Al , Ga, In, Tl ) by applying a hydrostatic pressure.

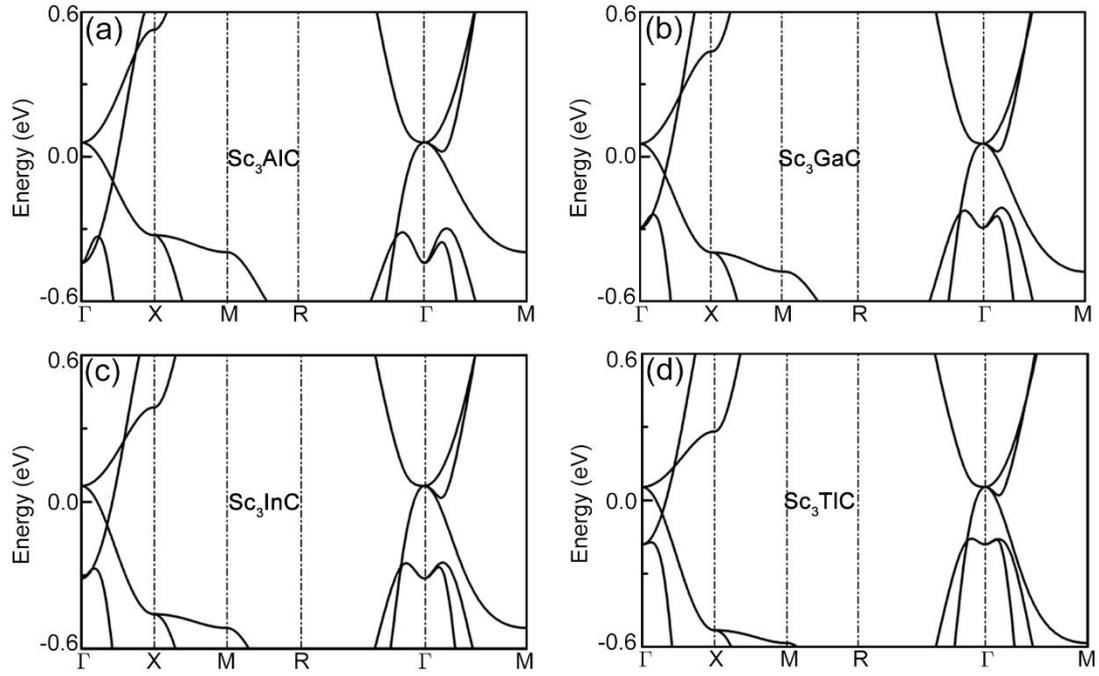

FIG. 8. (a)-(d) Electronic band structures of Sc$_3$XC ( X = Al , Ga, In, Tl ) in case of no SOC.

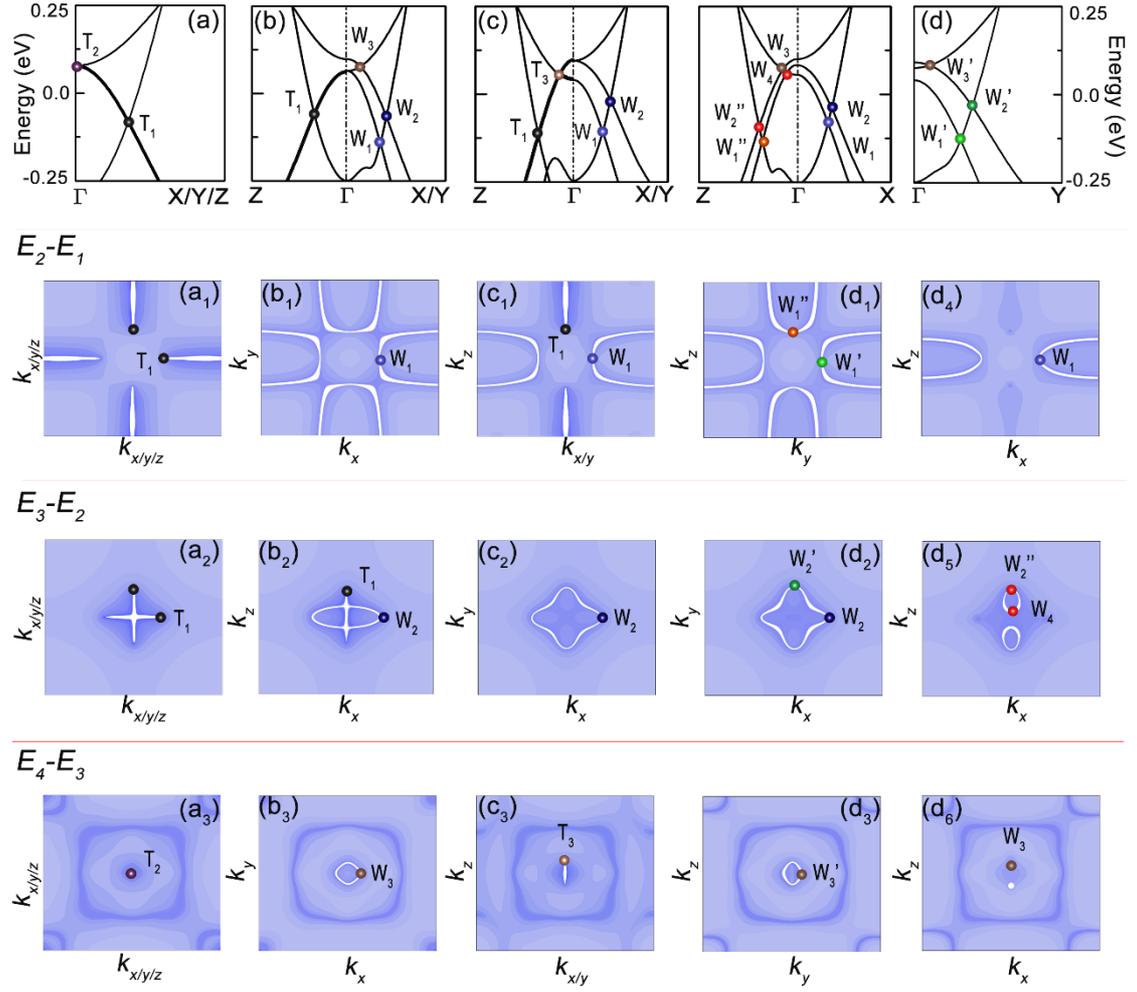

FIG. 9. DFT band structures for $Sc_3InC$ under different strains. (a) No strain. (b) 3% tensile strain along z axis. (c) 3% compressive strain along z axis. (d) 3% compressive strain along z axis and 1% compressive strain along y axis. The band structures in (a-d) corresponding to the band structures in Figs. 3(e-h), respectively. ($a_1$-$d_6$) Contour plots of energy difference between bands 1, 2, 3, 4 in (a-d). From these contour plots, the topological elements in the phases in Figs. 3(a-d) can be found clearly. For example, in ($a_1$-$a_2$), the triple points on axes are shown, which are connected by white straight nodal lines; in ($b_1$-$b_2$) or ($c_1$-$c_2$), the mutually vertical nodal rings of the 2D Hopf-chain network are shown.

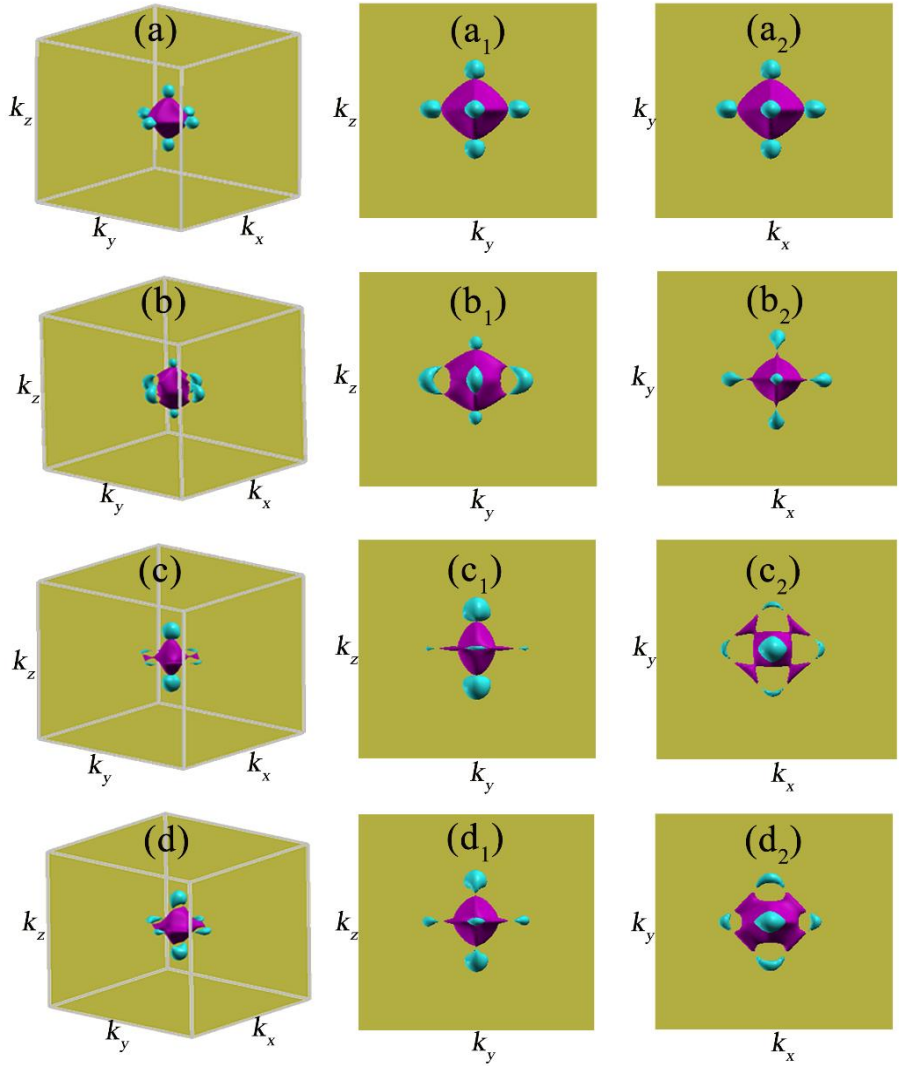

FIG. 10. (a-d) Fermi surfaces corresponding to the topological phases in Figs. 9(a-d), respectively. ($a_1$-$d_2$) show different side views of the Fermi surfaces.

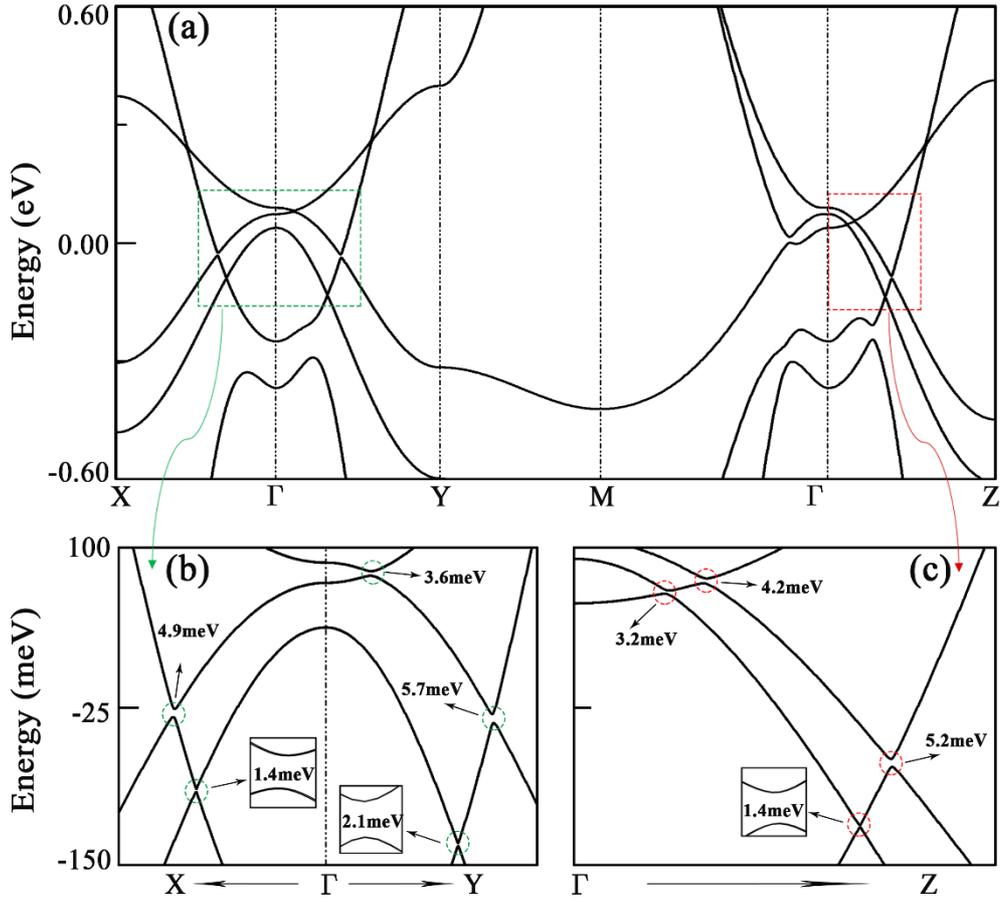

FIG. 11. (a) DFT band structure of Sc$_3$InC under a 3% compressive strain along z axis and a 1% compressive strain along y axis in the presence of SOC. (b) and (c) are amplified regions in the green and red dashed boxes in (a), respectively. The numbers in (b) and (c) show the widths of the gaps between energy bands. One can find that the gaps induced by SOC are very small (about 1~6 meV).

| Parameter | I | II | III | IV | V | VI | VII |
|---|---|---|---|---|---|---|---|
| $a_1$ | 0.08 | 0.06 | 0.10 | 0.11 | 0.08 | 0.10 | 0.11 |
| $a_2$ | 0.08 | 0.06 | 0.10 | 0.09 | 0.08 | 0.10 | 0.09 |
| $a_3$ | 0.08 | 0.10 | 0.06 | 0.06 | 0.08 | 0.06 | 0.06 |
| $a_4$ | −0.30 | −0.25 | −0.35 | −0.35 | 0.35 | −0.35 | −0.35 |
| $b_1$ | 0.09 | 0.07 | 0.08 | 0.07 | 0.13 | 0.10 | 0.11 |
| $b_2$ | −0.12 | −0.17 | −0.06 | −0.07 | −0.12 | −0.12 | −0.11 |
| $b_3$ | −0.12 | −0.10 | −0.10 | −0.09 | −0.12 | −0.15 | −0.15 |
| $b_4$ | 0.17 | 0.12 | 0.16 | 0.16 | 0.12 | 0.095 | 0.095 |
| $c_1$ | −0.12 | −0.17 | −0.06 | −0.06 | −0.12 | −0.12 | −0.13 |
| $c_2$ | 0.09 | 0.07 | 0.08 | 0.08 | 0.13 | 0.10 | 0.14 |
| $c_3$ | −0.12 | −0.10 | −0.10 | −0.10 | −0.12 | −0.15 | −0.15 |
| $c_4$ | 0.17 | 0.12 | 0.16 | 0.16 | 0.12 | 0.095 | 0.055 |
| $d_1$ | −0.12 | −0.12 | −0.09 | −0.08 | −0.12 | −0.05 | −0.03 |
| $d_2$ | −0.12 | −0.12 | −0.09 | −0.10 | −0.12 | −0.05 | −0.07 |
| $d_3$ | 0.09 | 0.07 | 0.08 | 0.08 | 0.13 | 0.08 | 0.08 |
| $d_4$ | 0.17 | 0.14 | 0.17 | 0.15 | 0.12 | 0.08 | 0.08 |
| A | 0.25 | 0.20 | 0.25 | 0.25 | 0.15 | 0.15 | 0.15 |
| B | 0.25 | 0.25 | 0.20 | 0.20 | 0.15 | 0.10 | 0.065 |
| C | −0.28 | −0.28 | −0.28 | −0.25 | −0.15 | −0.02 | −0.02 |
| D | 0.25 | 0.25 | 0.20 | 0.20 | 0.15 | 0.10 | 0.065 |
| E | −0.28 | −0.28 | −0.28 | −0.25 | −0.15 | −0.02 | −0.02 |
| F | −0.28 | −0.4 | −0.28 | −0.28 | −0.15 | −0.15 | −0.15 |

Table. 1. Detail values for the parameters in the k·p model in Eq. (1), which generate the topological phases in Figs. 3(a-d) and Figs. 7(a-c). The columns I-IV correspond to phases in Figs. 3(a-d) while V-VII correspond to those in Figs. 7(a-c), respectively.

| System | Sc$_3$AlC | Sc$_3$GaC | Sc$_3$InC | Sc$_3$TlC |
|---|---|---|---|---|
| Lattice parameter (Å) | 4.513 | 4.484 | 4.558 | 4.559 |
| D$_{Sc-C}$ (Å) | 2.256 | 2.242 | 2.279 | 2.279 |
| D$_{Sc-X}$ (Å) | 3.191 | 3.170 | 3.223 | 3.223 |
| D$_{C-X}$ (Å) | 3.908 | 3.883 | 3.947 | 3.948 |

Table. 2. Structure parameters of Sc$_3$XC (X = Al, Ga, In, Tl), in which D represents bond length between two types of atoms.